\documentstyle[preprint,aps]{revtex}

\draft

\begin{document}

\title{From quantum master equation to random walk}

\author{C. F. Huang}

\address{Department of Physics, National Taiwan University, Taipei,\\
Taiwan, R. O. C.}

\date{\today}

\maketitle

\begin{abstract}
It is shown in this paper that the quantum master equation can be mapped to
a modified continuous time random walk (CTRW) if the relaxation term is
composed of transitions over a set of states. When the Hamiltonian is
time-independent and transitions are between its eigenlevels, such a modified
CTRW reduces to the Markovian walk equivalent to the Pauli master equation.
On the other hand, the memory in such a modified CTRW is composed of a
temporal factor and the probability determined by the Liouville flow when
the relaxation term is reduced as a special dephasing term.
\end{abstract}


\newpage



\section{Introduction}

Random walks are important to the analysis in stochastic processes. [1-3]
Many physical phenomena can be mapped to random walks. It was shown that for
some nonMarkovian stochastic processes, we shall consider the continuous 
time random walk (CTRW). [1,4,5] In such a generalized random walk, the
probability density $r(l,t)$ for the walker to reach state $l$ at time $t$
is governed by
\begin{eqnarray}
r(l,t)=\sum_{l^{\prime }}\int_{t_{0}}^{t}dt^{\prime }\psi (l^{\prime
},l,t-t^{\prime })r(l^{\prime },t^{\prime })+s(l,t)
\end{eqnarray}
when $t\geq t_{0}$. Here the memory $\psi (l^{\prime },l,t-t^{\prime })$
presents the probability density to make a single transition to state $l$ at
time $t$ after reaching $l^{\prime }$ at time $t^{\prime }$, and the
nonnegative\ real function $s(l,t)$ presents the source term. In a CTRW, the
transition probability from $l^{\prime }$ to $l$ can depend on how long the
walker stays at $l^{\prime }$. The memory of the CTRW should satisfy
\begin{eqnarray}
\sum_{l}\int_{t^{\prime }}^{\infty }dt\psi (l^{\prime },l,t-t^{\prime })=1
\end{eqnarray}
so that the walker will make another transition after reaching any state $%
l^{\prime }$. For the CTRW starting at state $l_{0}$ at time $t=t_{i}$, we
shall set $s(l,t)=\delta _{l,l_{0}}\delta (t-t_{i})$ and $t_{0}=t_{i}$. The
Pauli master equation [6]
\begin{eqnarray}
\frac{d}{dt}p(l,t)=\sum_{l^{\prime }}w_{l^{\prime }l}p(l^{\prime
},t)-\sum_{l^{\prime }}w_{ll^{\prime }}p(l,t)
\end{eqnarray}
equivalent to a Markovian random walk, in fact, can be mapped to a CTRW with
\begin{eqnarray}
r(l,t)=\sum_{l^{\prime }}w_{l^{\prime }l}p(l^{\prime },t)\text{ and}
\end{eqnarray}
\[
\psi (l^{\prime },l,t-t^{\prime })=w_{l^{\prime }l}exp[-(t-t^{\prime
})\sum_{l^{\prime \prime }}w_{l^{\prime }l^{\prime \prime }}]
\]
if $\sum_{l^{\prime \prime }}w_{l^{\prime }l^{\prime \prime }}>0$ for all $%
l^{\prime }$, [1] as shown in Appendix A. Here $p(l,t)$ is the probability
to find the walker at $l$ at time $t$, and the nonnegative coefficient $%
w_{l^{\prime }l}$ is the rate for the transition $l^{\prime }\rightarrow l$.

On the other hand, for the time evolution of a quantum system described by
the (positive) density matrix $\rho (t)$ with
\begin{eqnarray}
tr\rho (t)=1,
\end{eqnarray}
it was shown that a (quantum) relaxation term should be included in addition
to the Liouville flow if the system is in contact with the reservoir. [7-14]
Let $H(t)$ be the hermitian operator for the Liouville flow. The density
matrix $\rho (t)$ is governed by
\begin{eqnarray}
\frac{\partial }{\partial t}\rho (t)=i[\rho (t),H(t)]+\frac{\partial }{%
\partial t}\rho (t)|_{relaxation},
\end{eqnarray}
where $\frac{\partial }{\partial t}\rho (t)|_{relaxation}$ is the (quantum)
relaxation term. When $\frac{\partial }{\partial t}\rho (t)|_{relaxation}=0$%
, the above equation is reduced to the (quantum) Liouville equation. [7] In
this paper, we take the constant $\hbar =1$, and define $[C,B]=CB-BC$ and $%
\{C,B\}=CB+BC$ for any two operators $B$ and $C$. If in Eq. (6) we set the
relaxation term as the Lindblad form [7,12] so that
\begin{eqnarray}
\frac{\partial }{\partial t}\rho (t)|_{relaxation}=\sum_{m}[-\frac{1}{2}%
\{\rho (t),{\cal V}_{m}{\cal V}_{m}^{\dagger }\}+{\cal V}_{m}^{\dagger }\rho
(t){\cal V}_{m}]
\end{eqnarray}
with $\{{\cal V}_{m}\}$ is a set of operators, the postivity of $\rho (t)$\
is preserved under Eq. (6). [7,8] If the
relaxation term is composed of transitions over a set $S$ in which each
state $n$ corresponds to a normalized ket $|n\rangle $, we shall set ${\cal V%
}_{m}=w_{nn^{\prime }}^{1/2}|n\rangle \langle n^{\prime }|$ for the
transition from $n$ to $n^{\prime }$ and replace the index $m$ by $%
(n,n^{\prime })$, where $w_{nn^{\prime }}>0$ presents the transition rate.
Then Eq. (6) becomes
\begin{eqnarray}
\frac{\partial }{\partial t}\rho (t)=i[\rho (t),H(t)]-\{\rho
(t),A\}+\sum_{nn^{\prime }}w_{nn^{\prime }}|n^{\prime }\rangle \langle
n|\rho (t)|n\rangle \langle n^{\prime }|,
\end{eqnarray}
where
\begin{eqnarray}
A=\sum_{nn^{\prime }}\frac{w_{nn^{\prime }}}{2}|n\rangle \langle n|
\end{eqnarray}
presents the imaginary part of the self energy. [10] In many cases, $%
H(t)=H_{0}+V(t)$ with $H_{0}$ and $V(t)$ as the unperturbed and the
perturbed part, and $S$ is composed of the eigenstates of $H_{0}$. [9-11]

In this paper, it is shown in section II that Eq. (8) can be mapped to a
modified CTRW over $S$ if we define
\begin{eqnarray}
r^{(q)}(n,t)=\sum_{n^{\prime }}w_{n^{\prime }n}\langle n^{\prime }|\rho
(t)|n^{\prime }\rangle
\end{eqnarray}
as the probability density for the walker to reach $n$ at time $t$. (It is
not necessary to set $S$ as an orthogonal or a complete set in section II.)
In such a modified CTRW, we indroduce the memory $\psi ^{(q)}(n^{\prime
},t^{\prime },n,t)$ for the probability density to make a single transition
to $n$ at time $t$ after reaching $n^{\prime }$ at time $t^{\prime }$. In
comparison with the memory of a CTRW, $\psi ^{(q)}(n^{\prime },t^{\prime
},n,t)$ depends on both $t$ and $t^{\prime }$ rather than $t-t^{\prime }$,
the time the walker wastes at the state $n^{\prime }$ before jumping to $n$.
If $H(t)=H_{0}$ as a time-independent Hamiltonian and $S$ is composed of the
eigenstates of $H_{0}$, after calculations we have
\begin{eqnarray}
\frac{\partial }{\partial t}\langle n|\rho (t)|n\rangle =\sum_{n^{\prime
}}[w_{n^{\prime }n}\langle n^{\prime }|\rho (t)|n^{\prime }\rangle
-w_{nn^{\prime }}\langle n|\rho (t)|n\rangle ]
\end{eqnarray}
from Eq. (8). Comparing with Eq.
(3), we can see that the above equation is just the Pauli master
equation after replacing $n$ and $\langle n|\rho (t)|n\rangle $ by $l$ and $%
p(l,t)$, respectively. Therefore, from Eq. (4) we can see that the definition
of $r^{(q)}(n,t) $ is natural. On the other hand, when $\frac{\partial }{%
\partial t}\rho (t)|_{relaxation}$ is reduced as a pure dephasing term under
which all phases have the lifetime $\tau $, the memory $\psi
^{(q)}(n^{\prime },t^{\prime },n,t)$\ of the modified CTRW becomes
\begin{eqnarray}
\frac{1}{\tau }e^{-(t-t^{\prime })/\tau }|\langle n|U(t,t^{\prime
})|n^{\prime }\rangle |^{2}.
\end{eqnarray}
Here the unitary operator $U(t,t^{\prime })$ satisfies $i\frac{\partial }{%
\partial t}U(t,t^{\prime })=H(t)U(t,t^{\prime })$ and $U(t^{\prime
},t^{\prime })=I$, the identity operator.\ The factor $|\langle
n|U(t,t^{\prime })|n^{\prime }\rangle |^{2}$, in fact, is just the
probability for being $|n^{\prime }\rangle $ at time $t^{\prime }$ to jump
to $|n\rangle $ at time $t$ under the Liouville flow. These two examples and
some questions are discussed in section III. Conclusions are made in section
IV.

\section{Quantum master equation and random walk}

To relate Eq. (8) to a modified CTRW, note that the density matrix $\rho (t)$
governed by Eq. (8) satisfies
\begin{eqnarray}
\rho (t)=\sum_{n}\int_{t_{0}}^{t}dt^{\prime }r^{(q)}(n,t^{\prime
})G(t,t^{\prime })|n\rangle \langle n|G^{\dagger }(t,t^{\prime
})+G(t,t_{0})\rho (t_{0})G^{\dagger }(t,t_{0}),
\end{eqnarray}
as shown in Appendix B. Here the operator $G(t,t^{\prime })$ satisfies
\begin{eqnarray}
i\frac{\partial }{\partial t}G(t,t^{\prime })=(H(t)-iA)G(t,t^{\prime })
\end{eqnarray}
when $t>t^{\prime }$ and $G(t^{\prime },t^{\prime })=I$. To prove Eq. (13),
in fact, we just need to take the time derivative on both sides of Eq. (13)
and check the initial condition because Eq. (8) is a first-order
differential equation with respect to $t$. Inserting Eq. (13) into the right
hand side of Eq. (10), we have
\begin{eqnarray}
r^{(q)}(n,t)=\sum_{n^{\prime }}\int_{t_{0}}^{t}dt^{\prime }\psi
^{(q)}(n^{\prime },t^{\prime },n,t)r^{(q)}(n^{\prime },t^{\prime
})+s^{(q)}(n,t)
\end{eqnarray}
if we define
\begin{eqnarray}
\psi ^{(q)}(n^{\prime },t^{\prime },n,t)\equiv \sum_{n^{\prime \prime
}}w_{n^{\prime \prime }n}|\langle n^{\prime \prime }|G(t,t^{\prime
})|n^{\prime }\rangle |^{2}
\end{eqnarray}
and
\begin{eqnarray}
s^{(q)}(n,t)\equiv \sum_{n^{\prime \prime }}w_{n^{\prime \prime }n}\langle
n^{\prime \prime }|G(t,t_{0})\rho (t_{0})G^{\dagger }(t,t_{0})|n^{\prime
\prime }\rangle .
\end{eqnarray}
In comparison with Eq. (1), Eq. (15) can be taken as the equation for the
modified CTRW equivalent to Eq. (8) if $s^{(q)}(n,t)$ and $\psi
^{(q)}(n^{\prime },t^{\prime },n,t)$ can serve as the source term and
memory, respectively. It is easy to see that $\psi ^{(q)}(n^{\prime
},t^{\prime },n,t)$, and $s^{(q)}(n,t)$ are nonnegative from their defintion.

To take $\psi ^{(q)}(n^{\prime },t^{\prime },n,t)$ as the memory for the
transition probability density, we need to prove that the probability to
make another transition after reaching $n^{\prime }$ at time $t^{\prime }$
cannot be larger than 1 for all $n^\prime$ and $t^\prime$. That is, $%
\sum_{n}\int_{t^{\prime }}^{\infty}dt\psi ^{(q)}(n^{\prime}, t^{\prime},n,t)
\leq 1$ for all $n^{\prime }$ and $t^{\prime}$. To prove it, note that
\begin{eqnarray}
\frac{\partial }{\partial t}Tr[G^{\dagger }(t,t^{\prime })G(t,t^{\prime
})Q]=-2Tr[QG^{\dagger }(t,t^{\prime})AG(t,t^{\prime})]
\end{eqnarray}
for any (time-independent) positive hermitian operator $Q$ whose trace is
well-defined. Because $\sum_{n}\psi ^{(q)}(n^{\prime },t^{\prime },n,t)=
2Tr[QG^{\dagger }(t,t^{\prime })AG(t,t^{\prime})]|_{ Q=|n^{\prime }\rangle
\langle n^{\prime }|}$ after some calculations, we have
\begin{eqnarray}
\lim_{T\rightarrow \infty }\sum_{n}\int_{t^{\prime }}^{T}dt\psi ^{(q)} (
n^{\prime },t^{\prime },n,t)= \lim_{T\rightarrow \infty }\int_{t
^{\prime } }^{T} 2 Tr [ Q G ^{\dagger}(t,t^{\prime})AG(t,t^{\prime})]|
_{Q=|n ^{\prime}\rangle \langle n^{\prime }|} dt
\end{eqnarray}
\[
=-\lim_{T\rightarrow \infty }\int_{t^{\prime } }^{T} \frac{\partial }{
\partial t}Tr[G^{\dagger }(t,t^{\prime })G(t,t^{\prime})Q]|_{ Q=|n^{\prime }
\rangle \langle n^{\prime }|}
\]
\[
=-\lim_{T\rightarrow \infty } Tr( G^{\dagger} (t,t^{\prime })G(t,t^{\prime})
|n^{\prime }\rangle \langle n^{\prime}|)|_{t=t^{\prime}}^{t=T}
\]
\[
=1-\lim_{T\rightarrow \infty}||G(T,t^{\prime})|n^{\prime}\rangle||^{2}\leq
1.
\]
Since $\psi ^{(q)}(n^{\prime },t^{\prime },n,t)\geq 0$, $\lim_{T\rightarrow
\infty }\sum_{n}\int_{t^{\prime }}^{T}dt\psi ^{(q)}(n^{\prime },t^{\prime
},n,t)$ is well-defined and we can take $%
1-\lim_{T\rightarrow \infty }\sum_{n}\int_{t^{\prime }}^{T}dt\psi
^{(q)}(n^{\prime },t^{\prime },n,t)$ as the probability to be trapped in $%
n^{\prime }$ after reaching it at time $t^{\prime }$.

For $s^{(q)}(n,t)$, we have $\sum_{n}s^{(q)}(n,t)=2Tr(QG^{%
\dagger}(t,t_{0})AG (t,t_{0}))|_{Q=\rho (t_{0})}$ after calculations and
hence from Eq. (18), it is easy to prove that $\sum_{n}\int_{t_{0}}^{%
\infty}dt s^{(q)} (n,t) \leq tr \rho (t_{0})=1$. We can take $s^{(q)}(n,t)$
as the probability density for the walker to start at state $n$ and at time $%
t$, and $1-\sum_{n} \int_{t_{0}}^{ \infty }dts^{(q)}(n,t)$ as the
probability for the walker not to walk over $S$.

From the above proof, therefore, we can map Eq. (8) to a modified CTRW.

\section{Discussions}

In the last section, we have mapped Eq. (8) to a modified CTRW. When $%
H(t)=H_{0}$ as a time-independent Hamiltonian and the set $S$ for
transitions are composed of the eigenstates of $H_{0}$, the memory $\psi
^{(q)}(n^{\prime },t^{\prime },n,t)$ defined in Eq. (16) can be reduced as $%
w_{n^{\prime }n}e^{-(t-t^{\prime })\sum_{n^{\prime \prime }}w_{n^{\prime}n
^{\prime \prime }}}$. The calculations are easy because now the states in $S$
can be used to diagonalize both $H_{0}$\ and $A$. Replacing $n$ by $l$, in 
such a case $\psi ^{(q)}$ is just the memory for the Pauli master equation 
as that defined in Eq. (4), which is reasonable because we
have the Pauli master equation from Eq. (8). 

On the other hand, when $w_{nn^{\prime}}\rightarrow \delta
_{n,n^{\prime}}/\tau$ and $S$ is an orthonormal complete set, the operator $%
A $ defined in Eq. (9) reduces to $\frac{I}{2\tau}$ because $%
\sum_{n}|n\rangle \langle n|=I$. Now $G( t,t^{\prime })=U(t,t^{\prime
})e^{-(t-t^{\prime })/2\tau }$ after calculations and hence $\psi
^{(q)}(n^{\prime },t^{\prime },n,t)$ is reduced as that in Eq. (12). The
factor $|\langle n|U(t,t^{\prime})|n^{\prime} \rangle |^2 $ in Eq. (12) is
just the probability for being $|n^{\prime}\rangle$ at time $t^{\prime}$ to
jump to $|n\rangle $ at time $t$ under the Liouville flow if at $t$ we perform
a (quantum) measurement [7] under which the system becomes a state in $S$.
But it should be emphasized that we are discussing a system in
contact with the reservoir and $\psi^{(q)}$ is not obtained directly from
assumptions about (quantum) measurements [7]. (In fact, suitable relaxation 
terms are used for $\lq\lq$continuous quantum measurements". [16,17])
The factor $\frac{1}{\tau}
e^{-(t-t^{\prime})/{\tau}}$ is induced by $A$ which presents the imaginary
part of the self energy. Since $\frac{\partial}{\partial t} \langle n| \rho
(t) |n \rangle |_{relaxation} =0$ in such a case, there is no transition
over $S$ due to this relaxation term. But for phases, i.e. the off-diagonal
terms of $\rho (t)$ in the expansion with respect to $S$, we have
\begin{eqnarray}
\frac{\partial }{\partial t}\langle n|\rho (t)|n^{\prime }\rangle
|_{relaxation }=-\frac{1}{\tau }\langle n|\rho (t) | n^{\prime} \rangle,
\end{eqnarray}
where $n\not=n^{\prime }$. Therefore, such a relaxation term induces the
decay of phases without inducing transitions and is a pure dephasing term
 [10]. This pure dephasing term is a special case of the relaxation term in
Eq. (7), so it will not break the positivity of $\rho (t)$.
[7,8] If we include an arbitrary pure dephasing term, however, the positivity
of $\rho (t)$ may be broken in general. [13,14]

It is proved that $\sum_{n}\int_{t^{\prime }}^{\infty }dt\psi
^{(q)}(n^{\prime },t^{\prime },n,t)\leq 1$ and $\sum_{n}\int_{t_{0}}^{\infty
}dts^{(q)}(n,t)\leq tr\rho (t_{0})=1$ in the last section. If $S$ is an
orthonormal complete set and there is a positive real number $c$ so that
\begin{eqnarray}
\sum_{n^{\prime }}w_{nn^{\prime }}\geq c
\end{eqnarray}
for all $n$, and we can prove that $\sum_{n}\int_{t^{\prime }}^{\infty }dt\psi
^{(q)}(n^{\prime },t^{\prime},n,t)=1$ and $\sum_{n}\int_{t_{0}}^{\infty }dt s
^{(q)}(n,t)=tr\rho (t_{0})=1$. This is because in Eq. (18),
\begin{eqnarray}
2Tr[QG^{\dagger }(t,t^{\prime })AG(t,t^{\prime })]
\end{eqnarray}
\[
\geq c\sum_{n}\langle n|G(t,t^{\prime })QG^{\dagger }(t,t^{\prime
})|n\rangle
\]
\[
=cTr[G^{\dagger }(t,t^{\prime })G(t,t^{\prime })Q]
\]
if Eq. (21) holds. From Eqs. (18) and (22) we have $\frac{\partial}{\partial t}
Tr[G^{\dagger }(t,t^{\prime })G(t,t^{\prime })Q]\leq -cTr[G^{\dagger}
(t,t^{\prime })G(t,t^{\prime })Q]$, and hence
\begin{eqnarray}
\lim_{t\rightarrow \infty }Tr[G^{\dagger }(t,t^{\prime })G(t,t^{\prime
})Q]\leq \lim_{t\rightarrow \infty }e^{-c(t-t^{\prime })}TrQ=0.
\end{eqnarray}
Inserting Eq. (23) into Eq. (19) with $Q=|n^{\prime }\rangle \langle
n^{\prime }|$, we can complete the proof for $\psi ^{(q)}$. The proof for $%
s^{(q)}$ is similar. Therefore, if Eq. (21) holds and $S$ is an orthonormal
complete set, there is no probability for the walker not to walk or not to
make the next transition. But $\psi ^{(q)}(n^{\prime },t^{\prime
},n,t)|_{n=n^{\prime }}$ may be not zero and hence the walker may stay at
the same state under a transition. The random walk with the memory defined
in Eq. (12) is equivalent to Eq. (8) when $w_{nn^{\prime }}\rightarrow
\delta _{n,n^{\prime }}/\tau $ and $S$ as an orthonormal complete set, and
we can set $c=1/ \tau$ for such a walk to satisfy Eq. (21).

For a quantum system composed of many identical noninteracting particles,
usually $\rho (t)$ presents the one-body density matrix. So we do not use
Eq. (5) to normalize $\rho (t)$, but set $tr \rho (t)$ as the number
of particles. For a dilute system, we can still use Eq. (8) to include
transitions in addition to the Liouville flow. [13,14] Because Eq. (8) is
linear, it is unimportant how to normalize $\rho (t)$ when we map Eq.
(8) to a modified CTRW. Now $r^{(q)}(n,t)\Delta t$ presents the expected
number of particles to enter the orbital $n$ in the time interval $(t,
t+\Delta t)$ as $\Delta t\rightarrow 0$. If the system is not dilute,
however, in general Eq. (8) should be modified, as shown in Appendix A.
The modified equation in Refs. 13 and 14, in fact, can be obtained by
substituting
\begin{eqnarray}
A^{\prime }(t)=\sum_{nn^{\prime }}\frac{w_{nn^{\prime }}}{2}[(1-\alpha 
\langle n^{\prime }|\rho (t)|n^{\prime }\rangle )|n\rangle \langle n|+\alpha
\langle n|\rho (t)|n\rangle |n^{\prime }\rangle \langle n^{\prime }|],
\end{eqnarray}
for $A$ in Eq. (8), where the constant $\alpha=1$ for fermions and 
$\alpha=-1$ for bosons. Then Eq. (8) becomes
\begin{eqnarray}
\frac{\partial }{\partial t}\rho (t) =i[\rho (t),H(t)]-\sum_{nn^{\prime }}
\frac{w_{nn^{\prime }}}{2} \{ \rho (t),(1-\alpha \langle n ^{\prime}
|\rho (t)|n^{\prime }\rangle )|n\rangle \langle n|+ \alpha \langle
n|\rho (t)|n\rangle |n^{\prime }\rangle \langle n^{\prime }|\}+
\end{eqnarray}
\[
\sum_{nn^{\prime }}w_{nn^{\prime }}|n^{\prime }\rangle \langle n|\rho
(t)|n\rangle \langle n^{\prime }|.
\]
As shown in Appendix C, the positivity of $\rho (t)$\ is still kept and for
fermions $\langle a|\rho (t)|a\rangle \leq 1$ for any normalized $|a\rangle $.
The above equation, however, is nonlinear and in general it is not expected
that we can map the above equation to a random walk. But $A^{\prime }(t)$ can
be reduced to $A$ when $w_{nn^{\prime }}=w_{n^{\prime }n}$ and we can still
map such a case to the modified CTRW.

\section{Conclusions}

In this paper, it is shown that the quantum master equation can be mapped to
a modified continuous time random walk if the relaxation term is composed of
transitions over a set of states. When the Hamiltonian for the Liouville
flow is time-independent and the set is composed of the eigenstates of this
Hamiltonian, the modified CTRW is just the Markovian walk corresponding to
the Pauli master equation. On the other hand, if the relaxation term is
reduced as a special pure dephasing term, the memory is composed of the
probability due to the Liouville flow and a temporal factor due to the
imaginary part of the self energy.

\section*{Acknowledgments}

The author thanks C. C. Chang for his valuable discussions.

\section*{Appendix A}

To prove that Eq. (3) can be mapped to a CTRW, note that the solution $y(t)$
of the differential equation
\begin{eqnarray}
\frac{d}{dt}y(t)=-\lambda y(t)+h(t) \text{ with } y(t_{0})=y_{0}
\end{eqnarray}
is of the form
\begin{eqnarray}
y(t)=\int_{t_{0}}^{t}dt^{\prime }e^{-\lambda (t-t^{\prime })}h(t^{\prime
})+e^{-\lambda (t-t_{0})}y_{0},
\end{eqnarray}
where $\lambda$ and $y_{0}$\ are real numbers, and $h(t)$ is a known real
function. If in Eq. (3) we set $p(l,t)$, $\sum_{l^{\prime }}w_{ll^{\prime }}$, 
$p(l,t_{0})$, and $r(l,t)(=\sum_{l^{\prime}}w_{l^{\prime }l}p(l^{\prime },t))$ 
as $y(t)$, $\lambda$, $y_{0}$, and $h(t)$, respectively, Eq. (3)
can be reduced as Eq. (26). Hence from Eq. (27), we have
\begin{eqnarray}
p(l,t)=\int_{t_{0}}^{t}dt^{\prime }e^{-(t-t^{\prime })\sum_{l^{\prime
}}w_{ll^{\prime }}}r(l,t^{\prime })+e^{-(t-t_{0})\sum_{l^{\prime
}}w_{ll^{\prime }}}p(l,t_{0})
\end{eqnarray}
Inserting the above equation into the right hand side of the first equation
in Eq. (4), we can complete the proof.

On the other hand, as mentioned in section III, in general Eq. (8) should be
modified when $\rho (t)$ presents the one-body density matrix of systems
composed of many identical noninteracting particles. 
To see this, consider the case that $H(t)=H_{0}$ as a time-independent
Hamiltonian and each $n$ in $S$ satisfies $H_{0} |n\rangle =E_{n} |n\rangle$. 
(Then $\{|n\rangle\}$ is an orthonormal complete set.) Expanding
$\rho (t)$ by states in $S$ in such a case, from Eq. (8) we can obtain Eq.
(11) for the diagonal terms. For the phases (off-diagonal terms), we have
\begin{eqnarray}
\frac{\partial }{\partial t}\langle n|\rho (t)|n^{\prime }\rangle =-\frac{1}{
2}\langle n|\rho (t)|n^{\prime }\rangle \sum_{n^{\prime \prime
}}(w_{nn^{\prime \prime }}+w_{n^{\prime }n^{\prime \prime }})+
i(E_{n^{\prime}}-E_{n}) \langle n | \rho (t)| n^{\prime} \rangle
\end{eqnarray}
for any two different (normalized)\ kets $|n\rangle $ and $|n^{\prime
}\rangle $ $\in $ $S$. (In the above equation, in fact, the first term at 
the right hand side describes the decoherence, i.e. the decay of phases.
[9-11]) Semiclassically, however, Eq. (3) should be modified as [13-15]
\begin{eqnarray}
\frac{\partial }{\partial t}p(l,t)=\sum_{l^{\prime }}[w_{l^{\prime}
l}(1-\alpha p(l,t))p(l^{\prime },t)
\end{eqnarray}
\[
-w_{l l^{\prime}}p(l,t)(1-\alpha p(l^{\prime },t))],
\]
which is not equivalent to Eq. (11) in general. Here the constant $\alpha$ 
equals 1 and -1 for fermions and bosons, respectively. In addition, for 
fermions Eq. (29) should also be modified to keep
\begin{eqnarray}
\langle a|\rho (t)|a\rangle \leq 1
\end{eqnarray}
for any normalized ket $|a\rangle$. [14] But if we modified Eq. (8) as Eq. 
(25), we have
\begin{eqnarray}
\frac{\partial }{\partial t}\langle n|\rho (t)|n\rangle =\sum_{n^{\prime
}}[w_{n^{\prime }n}(1-\alpha \langle n|\rho (t)|n\rangle )\langle n^{\prime
}|\rho (t)|n^{\prime }\rangle
\end{eqnarray}
\[
-w_{nn^{\prime }}\langle n|\rho (t)|n\rangle (1-\alpha \langle n^{\prime }|
\rho (t)|n^{\prime }\rangle )]
\]
when $H(t)=H_{0}$ and $S$ is composed of the eigenstates of $H_{0}$. The 
above equation is just Eq. (30) if we replace $\langle n|\rho (t)|n\rangle$ 
and $n$ by $p(l,t)$ and $l$, respectively. In addition, as shown in Appendix
C, for fermions Eq. (31) is kept under Eq. (25) for any arbitrary $H(t)$ and
$S$. Therefore, the modification is reasonable.

\section*{Appendix B}

Consider the time-dependent matrix $M(t)={\cal G} (t,t^{\prime}) M_{0} {\cal G}
^{\dagger} (t,t^{\prime })$ with the operator ${\cal G} (t,t^{\prime})$
satisfying $i\frac{\partial }{\partial t}{\cal G} (t,t^{\prime})=({\cal H}(t)
-i{\cal A}(t)){\cal G}(t,t^{\prime })$ at $t \geq t^{\prime}$ and ${\cal G}
(t^{\prime},t^{\prime })=I$, where ${\cal H}(t)$ and ${\cal A}(t)$ are two
time-dependent hermitian operators and $M_{0}$ is an arbitrary operator. When
$t^{\prime }$ is just the initial time $t_{0}$, the matrix $M(t)$ is the
unique one satisfying
\begin{eqnarray}
\frac{\partial }{\partial t}M(t)=i[M(t),{\cal H}(t)]-\{M(t),{\cal A}(t)\}
\end{eqnarray}
and $M(t_{0})=M_{0}$. If $t^{\prime }$ is larger than the initial time
$t_{0}$ and we assume that $M(t)=0$ when $t<t^{\prime}$,
the matrix $M(t)$ is the uique one satisfying $M(t_{0})=0$ and 
\begin{eqnarray}
\frac{\partial }{\partial t}M(t)=i[M(t),{\cal H}(t)]-\{ M (t),{\cal A}
(t)\}+{\cal S}(t)
\end{eqnarray}
with ${\cal S} (t)=M_{0}\delta (t-t^{\prime})$. Therefore, if ${\cal S}(t)=\sum
_{j} Q_{j} \delta (t-t_{j})$, it is not hard to see that the time-dependent
matrix $O(t)$ satisfying
\begin{eqnarray}
\frac{\partial }{\partial t}O(t)=i[O(t),{\cal H}(t)]-\{ O (t),{\cal A}%
(t)\}+{\cal S} (t)
\end{eqnarray}
with $O(t_{0})=O_{0}$ is
\begin{eqnarray}
O(t)=\sum_{t_{j}<t}{\cal G}(t,t_{j})Q_{j}{\cal G}^{\dagger }
(t,t_{j})+{\cal G}(t,t_{0})O_{0}{\cal G}^{\dagger }(t,t_{0})
\end{eqnarray}
Here $Q_{j}$ and $O_{0}$ are arbitrary operators. In the above two
equations, we shall replace $Q_{j}$ and $\sum_{t_{j}<t}$ by $dt^{\prime 
\prime} Q(t^{\prime \prime})$ and $\int_{t_{0}}^{t} $, respectively, and 
${\cal S} (t)=Q(t)$ in the continuous limit. Then we can see that the 
the solution of Eq. (8) can be written as that in Eq. (13) with ${\cal S}(t) 
\rightarrow \sum_{nn^{\prime }}w_{nn^{\prime }}|n^{\prime }\rangle \langle
n|\rho (t)|n\rangle \langle n^{\prime }|$, ${\cal H}(t) \rightarrow H(t)$, 
${\cal A} (t) \rightarrow A$, $O(t)\rightarrow \rho (t)$, $O_{0} \rightarrow
\rho (t_{0})$, and ${\cal G} (t,t^{\prime})\rightarrow G(t,t^{\prime})$.

\section{Appendix C}

To show that Eq. (25) is a suitable modification, we need to prove that
the positivity of $\rho (t)$ is still kept. That is, the equation
\begin{eqnarray}
\langle b | \rho (t)| b \rangle \geq 0,
\end{eqnarray}
is kept for any ket $| b \rangle$. In addition,
for any $H(t)$ and $S$, for fermions we need to prove that Eq. (31) holds
for any normalized $|a\rangle $. Eq, (31), in fact, is equivalent to
\begin{eqnarray}
\langle b|\rho (t)|b\rangle \leq \langle b|b\rangle
\end{eqnarray}
for any ket $|b\rangle $. The complete proof will be shown in Ref. 14. In
the following I just show the key points. Redefining $G(t,t^{\prime}
)$ by substituting $A^{\prime }(t)$ for $A$ in Eq. (14), under Eq. (25)
first we still have Eq. (13)
\[
\rho (t)=\sum_{n}\int_{t_{0}}^{t}dt^{\prime }r^{(q)}(n,t^{\prime
})G(t,t^{\prime })|n\rangle \langle n|G^{\dagger }(t,t^{\prime
})+G(t,t_{0})\rho (t_{0})G^{\dagger }(t,t_{0}).
\]
Secondly, setting $\alpha=1$ for fermions, we have
\begin{eqnarray}
\rho (t)=I-\sum_{n}\int_{t_{0}}^{t}dt^{\prime }R(n,t^{\prime })G(t,t^{\prime
})|n\rangle \langle n|G^{\dagger }(t,t^{\prime })
\end{eqnarray}
\[
-G(t,t_{0})G^{\dagger }(t,t_{0})+G(t,t_{0})\rho (t_{0})G^{\dagger
}(t,t_{0}),
\]
where $R(n,t^{\prime })\equiv \sum_{n^{\prime }}w_{nn^{\prime}}(1- \langle
n^{\prime }|\rho (t)|n^{\prime }\rangle )$. To prove the above two equations,
we just need to check the initial condition and their time derivatives.
Assume that $\rho $ is positive at any time 
$t^{\prime }<t$. If $\alpha =1$, we also assume that $\langle b|\rho (t
^{\prime} )|b \rangle \leq \langle b|b\rangle$ for any ket $|b\rangle$
when $t^{\prime }<t$. Then at time $t$, for any normalized ket $|m\rangle$
we have
\begin{eqnarray}
\rho (t)=\sum_{n}\int_{t_{0}}^{t}dt^{\prime }r^{(q)}(n,t^{\prime })|\langle
\ell (m,t,t^{\prime })|n\rangle |^{2}
\end{eqnarray}
\[
+\langle \ell (m,t,t_{0})|\rho (t_{0})|\ell (m,t,t_{0})\rangle \geq 0,
\]
from Eq. (13), where the ket $\ell (m,t,t^{\prime })\equiv G^{\dagger }(t,t^{
\prime})|m\rangle $ for any $|m\rangle $. In addition, for fermions,
\begin{eqnarray}
\langle m|\rho (t)|m\rangle =\langle m|m\rangle
-\sum_{n}\int_{t_{0}}^{t}dt^{\prime }R(n,t^{\prime })|\langle \ell
(m,t,t^{\prime })|n\rangle |^{2}
\end{eqnarray}
\[
-[\langle \ell (m,t,t_{0})|\ell (m,t,t_{0})\rangle -\langle \ell
(m,t,t_{0})|\rho (t_{0})|\ell (m,t,t_{0})\rangle ]\leq \langle m|m\rangle
\]
from Eq. (39). We can see from Eqs. (40) and (41) that under Eq. (25), Eq. (38)
is kept for fermions and the positivity of $\rho (t)$ is kept for fermions
and bosons.

\end{document}